\documentclass[aps,showpacs,tightenlines,nobalancelastpage,
floatfix,onecolumn]{revtex4}

\usepackage{graphicx}
\usepackage{amsmath}
\usepackage{amsfonts}
\usepackage{amssymb}
\usepackage{bbm}
\usepackage{epic}

\begin{document}

\title{Stimulated Raman adiabatic passage in a multi-level atom}

\date{\today}

\author{A. D. Boozer} 

\affiliation{
  Norman Bridge Laboratory of Physics 12-33,
  California Institute of Technology,
  Pasadena, CA 91125
}

\begin{abstract}
  We present a general formalism for describing stimulated Raman
  adiabatic passage in a multi-level atom.
  The atom is assumed to have two ground state manifolds $a$ and $b$
  and an excited state manifold $e$, and the adiabatic
  passage is carried out by resonantly driving the $a-e$ and $b-e$
  transitions with time-dependent fields.
  Our formalism gives a complete description of the adiabatic passage
  process, and can be applied to systems with
  arbitrary numbers of degenerate states in each manifold and
  arbitrary couplings of the $a-e$ and $b-e$ transitions.
  We illustrate the formalism by applying it to both a simple toy
  model and to adiabatic passage in the Cesium atom.
\end{abstract}


\maketitle

\section{Introduction}

A number of techniques for coherently manipulating atomic systems are
based on the idea of adiabatically varying the parameters of a
time-dependent Hamiltonian.
One such example is the technique of stimulated Raman
adiabatic passage (STIRAP), which can be used to coherently transfer
population between the two ground states of a three-level atom in the
lambda configuration \cite{hioe83, oreg84, kuklinski89}.
The STIRAP technique involves driving the atom with a pair of
time-dependent fields that couple the two ground states to the excited
state, and relies on the fact that the Hamiltonian for the system has
an instantaneous eigenstate, called the dark state, that contains no
excited state component.
In general this dark state is a superposition of the two
ground states, where the overlap of the dark state with each ground
state depends on the ratio of the powers in the two fields.
Population is transfered from the first ground state to the second
by adiabatically varying these powers in such a way that initially the
dark state overlaps entirely with the first ground state, during the
process the dark state is a superposition of the two ground states,
and after the process has been completed the dark state overlaps
entirely with the second ground state.
The STIRAP technique has been studied experimentally
\cite{gaubatz88, gaubatz90, rubahn91, broers92, schiemann93,
  weitz94, boozer07},
and has formed the basis for numerous theoretical proposals
\cite{marte91, unanyan98, theuer99, unanyan99}.
In addition, STIRAP has been generalized to multi-level systems
\cite{shore95, martin95, kis01, kiz04}, opening up new possibilities
for creating superposition states
\cite{chang01, kis02, karpati03, kis05,thanopulos}.

Here we present a general formalism for describing STIRAP in a
multi-level atom.
The atom is assumed to have two ground state
manifolds $a$ and $b$ and an excited state manifold $e$, where each
manifold consists of an arbitrary number of degenerate states,
and the adiabatic passage is carried out by resonantly driving the
$a-e$ and $b-e$ transitions with time-dependent fields.
The Hamiltonian for the system is
\begin{eqnarray}
  H =
  \frac{\Omega}{2}\cos\theta \,(A_b + A_b^\dagger) -
  \frac{\Omega}{2}\sin\theta \,(A_a + A_a^\dagger),
\end{eqnarray}
where $\Omega$ is a time-independent Rabi frequency that characterizes
the strength of the fields, $\theta$ is a time-dependent parameter
that is varied to carry out the adiabatic passage, and
$A_a$ and $A_b$ are lowering operators that connect
states in manifold $e$ to states in manifolds $a$ and $b$.
We sweep $\theta$ from $0$ to $\pi/2$ to transfer states
from manifold $a$ to manifold $b$, and we sweep $\theta$ from $\pi/2$
to $0$ to transfer states from manifold $b$ to manifold $a$.

Let us denote the Hilbert spaces for manifolds $a$, $b$, and $e$ by
${\cal H}_a$, ${\cal H}_b$, and ${\cal H}_e$.
We will show that ${\cal H}_a$ and ${\cal H}_b$ can
be decomposed as follows:
\begin{eqnarray}
  \label{eqn:decomposition-a}
  {\cal H}_a & = &
  {\cal H}_a^\lambda \oplus {\cal H}_a^d \oplus {\cal H}_a' \\
  \label{eqn:decomposition-b}
  {\cal H}_b & = &
  {\cal H}_b^\lambda \oplus {\cal H}_b^d \oplus {\cal H}_b'.
\end{eqnarray}
Under adiabatic passage from $a$ to $b$, states in
${\cal H}_a^\lambda$ coherently evolve into states in
${\cal H}_b^\lambda$, states in ${\cal H}_a^d$ remain unchanged, and
states in ${\cal H}_a'$ are driven to the excited state manifold and
scatter incoherently.
Similarly, for adiabatic passage from $b$ to $a$, states in
${\cal H}_b^\lambda$ coherently evolve into states in
${\cal H}_a^\lambda$, states in ${\cal H}_b^d$ remain unchanged, and
states in ${\cal H}_b'$ are driven to the excited state manifold and
scatter incoherently.
The coherent evolution from ${\cal H}_a^\lambda$ to
${\cal H}_b^\lambda$ is described by a unitary transformation
$U:{\cal H}_a^\lambda \rightarrow {\cal H}_b^\lambda$, and
the coherent evolution from ${\cal H}_b^\lambda$ to
${\cal H}_a^\lambda$ is described by the unitary transformation
$U^\dagger:{\cal H}_b^\lambda \rightarrow {\cal H}_a^\lambda$.
Given arbitrary operators $A_a$ and $A_b$,
our goal is to perform the Hilbert space decompositions described
in equations (\ref{eqn:decomposition-a}) and
(\ref{eqn:decomposition-b}) and to calculate the unitary
transformation $U$.

The paper is organized as follows.
In section \ref{sec:decompositions}, we show how to preform the
Hilbert space decompositions described in equations
(\ref{eqn:decomposition-a}) and (\ref{eqn:decomposition-b}).
In section \ref{sec:unitary-transformation}, we calculate the unitary
transformation
$U:{\cal H}_a^\lambda \rightarrow {\cal H}_b^\lambda$, and consider in
detail the special case
$\dim {\cal H}_a^\lambda = \dim {\cal H}_b^\lambda = 2$.
The formalism developed in sections \ref{sec:decompositions} and
\ref{sec:unitary-transformation} gives a complete description of the
adiabatic transfer process, and in
section \ref{sec:example-system} we illustrate this formalism by
applying it to a simple toy model.
Finally, in section \ref{sec:cesium} we use the formalism to analyze
adiabatic passage in the Cesium atom.

\section{Hilbert space decompositions}
\label{sec:decompositions}

We will first show how to perform the Hilbert space decompositions
described in equations (\ref{eqn:decomposition-a}) and
(\ref{eqn:decomposition-b}) for a given pair of atomic lowering
operators $A_a$ and $A_b$.
Let ${\cal H}_a^d$ be the space of states in manifold $a$ that
are dark to the $a \rightarrow e$ transition,
let ${\cal H}_b^d$ be the space of states in manifold $b$ that
are dark to the $b \rightarrow e$ transition, and
let ${\cal H}_e^d$ be the space of states in manifold $e$ that
are dark to the $e \rightarrow b$ transition.
Note that ${\cal H}_a^d$, ${\cal H}_b^d$, and ${\cal H}_e^d$ are just
the null spaces of the operators
$A_a^\dagger$, $A_b^\dagger$, and $A_b$:
\begin{eqnarray}
  {\cal H}_a^d & = & 
  \{ |\psi_a\rangle \in {\cal H}_a \mid
  A_a^\dagger|\psi_a\rangle = 0 \} \\
  {\cal H}_b^d & = & 
  \{ |\psi_b\rangle \in {\cal H}_b \mid
  A_b^\dagger|\psi_b\rangle = 0 \} \\
  {\cal H}_e^d & = & 
  \{ |\psi_e\rangle \in {\cal H}_e \mid A_b|\psi_e\rangle = 0 \}.
\end{eqnarray}
Define ${\cal H}_a^\perp$, ${\cal H}_b^\perp$, and ${\cal H}_e^\perp$
to be the complements of these spaces in
${\cal H}_a$, ${\cal H}_b$, and ${\cal H}_e$:
\begin{eqnarray}
  {\cal H}_a & = & 
  {\cal H}_a^\perp \oplus {\cal H}_a^d \\
  {\cal H}_b & = & 
  {\cal H}_b^\perp \oplus {\cal H}_b^d \\
  {\cal H}_e & = & 
  {\cal H}_e^\perp \oplus {\cal H}_e^d.
\end{eqnarray}
Clearly, states in ${\cal H}_a^d$ and ${\cal H}_b^d$ are dark states
of $H$.
In addition, $H$ has dark states of the form
\begin{eqnarray}
  \label{eqn:lambda-dark-state}
  |\Lambda\rangle = \cos\phi\,|\psi_a\rangle +
  \sin\phi\,|\psi_b\rangle,
\end{eqnarray}
where $|\psi_a\rangle \in {\cal H}_a^\perp$ and
$|\psi_b\rangle \in {\cal H}_b^\perp$.
We will call such states lambda dark states, and say that
$|\psi_a\rangle$ and $|\psi_b\rangle$ form a lambda pair.
Since $H|\Lambda\rangle = 0$, we have that
\begin{eqnarray}
  \cos\theta\sin\phi\,A_b^\dagger |\psi_b\rangle =
  \sin\theta \cos\phi\,A_a^\dagger\,|\psi_a\rangle.
\end{eqnarray}
This equation must hold for all values of $\theta$, so $\phi$ must be
related to $\theta$ by
\begin{eqnarray}
  \label{eqn:angle-relation}
  \tan \phi = \lambda \tan\theta
\end{eqnarray}
for some value $\lambda$.
Thus, states $|\psi_a\rangle$ and $|\psi_b\rangle$ are related by
\begin{eqnarray}
  \label{eqn:pair-relation}
  \lambda A_b^\dagger |\psi_b\rangle =
  A_a^\dagger |\psi_a\rangle.
\end{eqnarray}
Note that by using equation (\ref{eqn:angle-relation}),
we can also express $|\Lambda\rangle$ as
\begin{eqnarray}
  |\Lambda\rangle = (\cos^2\theta + \lambda^2 \sin^2\theta)^{-1/2}\,
  (\cos\theta\,|\psi_a\rangle + \lambda\sin\theta\,|\psi_b\rangle).
\end{eqnarray}
Define ${\cal H}_a^\lambda$ to be the space of states
$|\psi_a\rangle \in {\cal H}_a^\perp$ such that
$A_a^\dagger |\psi_a\rangle \in {\cal H}_e^\perp$, and define
${\cal H}_a'$ to be the complement of ${\cal H}_a^\lambda$ in
${\cal H}_a^\perp$:
\begin{eqnarray}
  {\cal H}_a^\perp & = &
  {\cal H}_a^\lambda \oplus {\cal H}_a'.
\end{eqnarray}
Note that from the definitions of ${\cal H}_a'$ and ${\cal H}_e^d$, it
follows that $\dim {\cal H}_a' \le \dim {\cal H}_e^d$.

We will now show that for every normalized state
$|\psi_a\rangle \in {\cal H}_a^\lambda$, we can construct a normalized
state $|\psi_b\rangle \in {\cal H}_b^\perp$ such that
$|\psi_a\rangle$ and $|\psi_b\rangle$ form a lambda pair.
First, note that we can view the raising operator $A_a^\dagger$ as a
mapping $A_a^\dagger:{\cal H}_a \rightarrow {\cal H}_e$.
Since the image of ${\cal H}_a^\lambda$ under $A_a^\dagger$ lies
entirely in ${\cal H}_e^\perp$, we can define a new mapping
$R_a^\dagger:{\cal H}_a^\lambda \rightarrow {\cal H}_e^\perp$ by
\begin{eqnarray}
  R_a^\dagger|\psi_a\rangle \equiv
  A_a^\dagger|\psi_a\rangle,
\end{eqnarray}
where $|\psi_a\rangle$ is an arbitrary state in
${\cal H}_a^\lambda$.
Similarly, given the definitions of ${\cal H}_b^\perp$ and
${\cal H}_e^\perp$, we can define mappings
$R_b^\dagger:{\cal H}_b^\perp \rightarrow {\cal H}_e^\perp$
and
$R_b:{\cal H}_e^\perp \rightarrow {\cal H}_b^\perp$
by
\begin{eqnarray}
  R_b^\dagger|\psi_b\rangle & \equiv &
  A_b^\dagger|\psi_b\rangle \\
  R_b|\psi_e\rangle & \equiv &
  A_b|\psi_e\rangle,
\end{eqnarray}
where $|\psi_b\rangle$ and $|\psi_e\rangle$ are arbitrary states in
${\cal H}_b^\perp$ and ${\cal H}_e^\perp$.
Note that
the only state in ${\cal H}_a^\lambda$ that lies in the null space of
$A_a^\dagger$ is the null state,
the only state in ${\cal H}_b^\perp$ that lies in the null space of
$A_b^\dagger$ is the null state, and
the only state in ${\cal H}_e^\perp$ that lies in null space of $A_b$
is the null state; thus,
the null spaces of $R_a^\dagger$, $R_b^\dagger$, and $R_b$ are
trivial.
This means that the mapping
$R_b^\dagger R_b: {\cal H}_e^\perp \rightarrow {\cal H}_e^\perp$ is
invertible, and we can define a mapping
$M:{\cal H}_a^\lambda \rightarrow {\cal H}_b^\perp$ by
\begin{eqnarray}
  \label{eqn:mapping-M}
  M = R_b\,(R_b^\dagger R_b)^{-1}\,R_a^\dagger.
\end{eqnarray}
Define ${\cal H}_b^\lambda$ to be the image of $M$, and define
${\cal H}_b'$ to be the complement of ${\cal H}_b^\lambda$ in
${\cal H}_b^\perp$:
\begin{eqnarray}
  {\cal H}_b^\perp = {\cal H}_b^\lambda \oplus {\cal H}_b'.
\end{eqnarray}
Note that because the null spaces of
$R_a^\dagger$, $R_b^\dagger$, and $R_b$ are all trivial, the null
space of $M$ is also trivial, and therefore
$\dim {\cal H}_b^\lambda = \dim {\cal H}_a^\lambda$.
Given a normalized state $|\psi_a\rangle \in {\cal H}_a^\lambda$,
define $\lambda$ by
\begin{eqnarray}
  \lambda = \langle \psi_a | M^\dagger M|\psi_a\rangle^{1/2},
\end{eqnarray}
and define a normalized state $|\psi_b\rangle \in {\cal H}_b^\lambda$
by
\begin{eqnarray}
  \label{eqn:map-to-lambda-pair}
  |\psi_b\rangle = \frac{1}{\lambda} M|\psi_a\rangle.
\end{eqnarray}
Using equations (\ref{eqn:mapping-M}) and
(\ref{eqn:map-to-lambda-pair}), we find that 
\begin{eqnarray}
  \lambda A_b^\dagger |\psi_b\rangle =
  \lambda R_b^\dagger |\psi_b\rangle =
  R_b^\dagger M |\psi_a\rangle =
  R_b^\dagger R_b\,(R_b^\dagger R_b)^{-1}\,R_a^\dagger |\psi_a\rangle =
  R_a^\dagger|\psi_a\rangle =
  A_a^\dagger|\psi_a\rangle.
\end{eqnarray}
Thus, states $|\psi_b\rangle$ and $|\psi_a\rangle$ satisfy equation
(\ref{eqn:pair-relation}) and therefore form a lambda pair.

We claim that none of the states in ${\cal H}_a'$ can form lambda
pairs with states in ${\cal H}_b^\perp$.
To see this, consider a superposition of states
$|\psi_a\rangle \in {\cal H}_a'$ and
$|\psi_b\rangle \in {\cal H}_b^\perp$:
\begin{eqnarray}
  |\phi\rangle = c_a |\psi_a\rangle + c_b |\psi_b\rangle.
\end{eqnarray}
Because of the way we have defined ${\cal H}_a'$, there must be a
state $|\psi_e\rangle \in {\cal H}_e^d$ such that
\begin{eqnarray}
  \langle \psi_e |A_a^\dagger |\psi_a\rangle \neq 0.
\end{eqnarray}
Since $|\psi_e\rangle \in {\cal H}_e^d$ we have that
$A_b|\psi_e\rangle = 0$, so
\begin{eqnarray}
  \langle \psi_e |A_b^\dagger |\psi_b\rangle = 0,
\end{eqnarray}
and therefore
\begin{eqnarray}
  \langle \psi_e | H |\phi\rangle =
  -\frac{\Omega}{2} \sin\theta\,
  \langle \psi_e |A_a^\dagger | \psi_a\rangle\,c_a .
\end{eqnarray}
Thus, for $c_a \neq 0$ and $\sin \theta \neq 0$ we have that
$H|\phi\rangle \neq 0$, so $|\phi\rangle$ cannot be a lambda
dark state.

Since all the states in ${\cal H}_a^\lambda$ form lambda pairs with
states in ${\cal H}_b^\lambda$, and none of the states in
${\cal H}_a'$ form lambda pairs with states in ${\cal H}_b^\perp$, it
follows that none of the states in ${\cal H}_b'$ form lambda pairs
with states in ${\cal H}_a^\perp$.
Thus, states in ${\cal H}_a'$ and ${\cal H}_b'$ do not form dark
states of $H$, and under adiabatic passage they are driven to the
excited state manifold and scatter incoherently.

\section{Unitary transformation}
\label{sec:unitary-transformation}

In the previous section we defined Hilbert spaces ${\cal H}_a^\lambda$
and ${\cal H}_b^\lambda$, which are subspaces of the total Hilbert
spaces for ground state manifolds $a$ and $b$.
We will now show that under adiabatic passage from $a$ to $b$
states in ${\cal H}_a^\lambda$ coherently evolve
into states in ${\cal H}_b^\lambda$, and we will derive the unitary
transformation
$U:{\cal H}_a^\lambda \rightarrow {\cal H}_b^\lambda$ that describes
this evolution.

First, choose an orthonormal basis
$\{ |\psi_{a1}\rangle,\, \cdots,\, |\psi_{an}\rangle \}$
for ${\cal H}_a^\lambda$, and use the mapping $M$ given in equation
(\ref{eqn:mapping-M}) to construct a normalized basis
$\{ |\psi_{b1}\rangle,\, \cdots,\, |\psi_{bn}\rangle \}$
for ${\cal H}_b^\lambda$, where $|\psi_{bj}\rangle$ is defined by
\begin{eqnarray}
  \label{eqn:b-basis-states}
  |\psi_{bj}\rangle & = & \frac{1}{\lambda_j} M |\psi_{aj}\rangle \\
  \label{eqn:lambda-factors}
  \lambda_j & = &
  \langle \psi_{aj} | M^\dagger M|\psi_{aj} \rangle^{1/2}.
\end{eqnarray}
Note that the basis states for ${\cal H}_b^\lambda$ will not
necessarily be mutually orthogonal.
The states $|\psi_{aj}\rangle$ and $|\psi_{bj}\rangle$ form a lambda
pair, and define a lambda dark state that is given by
\begin{eqnarray}
  |\Lambda_j(\theta)\rangle =
  \cos\phi_j(\theta)\,|\psi_{aj}\rangle +
  \sin\phi_j(\theta)\,|\psi_{bj}\rangle,
\end{eqnarray}
where
\begin{eqnarray}
  \label{eqn:phi-j}
  \phi_j(\theta) = \tan^{-1}(\lambda_j\tan\theta).
\end{eqnarray}
The states
$\{ |\Lambda_1\rangle,\, \cdots,\, |\Lambda_n\rangle \}$
form a basis for the lambda dark states, so we can express a general
lambda dark state $|\Lambda\rangle$ as
\begin{eqnarray}
  |\Lambda\rangle = \sum_j c_j\,|\Lambda_j\rangle.
\end{eqnarray}
In the adiabatic limit, the time evolution of $|\Lambda\rangle$ is
given by Schr\"{o}dinger's equation:
\begin{eqnarray}
  i \frac{d}{dt} |\Lambda\rangle = H |\Lambda\rangle = 0.
\end{eqnarray}
Thus, substituting for $|\Lambda\rangle$, we find that
\begin{eqnarray}
  \label{eqn:schrodinger-dark}
  \sum_j (\dot{c}_j\,|\Lambda_j\rangle +
  c_j\,\dot{\phi}_j\,|\bar{\Lambda}_j\rangle) = 0,
\end{eqnarray}
where the dots indicate derivatives with respect to $\theta$, and
where we have defined
\begin{eqnarray}
  |\bar{\Lambda}_j\rangle = \frac{d}{d\phi_j}|\Lambda_j\rangle =
  -\sin\phi_j\,|\psi_{aj}\rangle +
  \cos\phi_j\,|\psi_{bj}\rangle.
\end{eqnarray}
Taking the inner product of equation (\ref{eqn:schrodinger-dark}) with
$\langle \Lambda_k|$, we obtain
\begin{eqnarray}
  \label{eqn:eqns-of-motion-temp}
  \sum_{j} (\dot{c}_j\,\langle \Lambda_k|\Lambda_j\rangle +
  c_j\,\dot{\phi}_j\,\langle \Lambda_k|\bar{\Lambda}_j\rangle) = 0.
\end{eqnarray}
Since 
$\{ |\Lambda_1\rangle,\, \cdots,\, |\Lambda_n\rangle \}$
is a complete basis for the lambda dark states, the matrix $\langle
\Lambda_k|\Lambda_j\rangle$ is invertible; we will denote its
inverse by $L_{ik}$:
\begin{eqnarray}
  \sum_k L_{ik}\,\langle \Lambda_k|\Lambda_j\rangle =
  \delta_{ij}.
\end{eqnarray}
If we multiply equation (\ref{eqn:eqns-of-motion-temp}) by $L_{ik}$
and then sum over $k$, we obtain the following equations of motion for
the expansion coefficients $c_i$:
\begin{eqnarray}
  \label{eqn:eqns-of-motion}
  \dot{c}_i =
  - \sum_j \sum_k L_{ik}\,\langle\Lambda_k|\bar{\Lambda}_j\rangle\,
  \dot{\phi}_j\,c_j.
\end{eqnarray}
Note that we can express an arbitrary state
$|\psi_a\rangle \in {\cal H}_a^\lambda$ as
\begin{eqnarray}
  |\psi_a\rangle =
  \sum_j c_j(0)\,|\psi_{aj}\rangle =
  \sum_j c_j(0)\,|\Lambda_j(0)\rangle.
\end{eqnarray}
for some set of amplitudes $c_j(0)$.
We can then integrate equation (\ref{eqn:eqns-of-motion}) subject to
these initial conditions to obtain a state
\begin{eqnarray}
  U|\psi_a\rangle =
  \sum_j c_j(\pi/2)\,|\Lambda_j(\pi/2)\rangle =
  \sum_j c_j(\pi/2)\,|\psi_{bj}\rangle.
\end{eqnarray}
This defines the unitary transformation $U$.

\subsection{Example:
  $\dim {\cal H}_a^\lambda = \dim {\cal H}_b^\lambda = 2$}
\label{sec:unitary-transformation-two-level}

As an example, we will  write down the equations of motion explicitly
for the case $\dim {\cal H}_a^\lambda = \dim {\cal H}_b^\lambda = 2$.
We choose an orthonormal basis of states
$\{ |\psi_{a1}\rangle,\, |\psi_{a2}\rangle \}$ for
${\cal H}_a^\lambda$,
use equation (\ref{eqn:b-basis-states}) to obtain a basis of states
$\{ |\psi_{b1}\rangle,\, |\psi_{b2}\rangle \}$ for
${\cal H}_b^\lambda$,
and use equation (\ref{eqn:lambda-factors}) to obtain the values
$\lambda_1$ and $\lambda_2$.
The matrices $\langle\Lambda_k|\Lambda_j\rangle$ and
$\langle\Lambda_k|\bar{\Lambda}_j\rangle$ are given by
\begin{eqnarray}
  \langle\Lambda_k|\Lambda_j\rangle & = &
  \left(
  \begin{array}{cc}
    1 & z \sin\phi_1\sin\phi_2 \\
    z^* \sin\phi_1\sin\phi_2 & 1
  \end{array}
  \right) \\
  \langle\Lambda_k|\bar{\Lambda}_j\rangle & = &
  \left(
  \begin{array}{cc}
    0 & z \sin\phi_1\cos\phi_2 \\
    z^* \cos\phi_1\sin\phi_2 & 0
  \end{array}
  \right),
\end{eqnarray}
where
\begin{eqnarray}
  z = \langle \psi_{b1} | \psi_{b2} \rangle.
\end{eqnarray}
The inverse of $\langle\Lambda_k|\Lambda_j\rangle$ is
\begin{eqnarray}
  L_{ik} & = &
  (1 - |z|^2 \sin^2\phi_1\sin^2\phi_2)^{-1}
  \left(
  \begin{array}{cc}
    1 & -z \sin\phi_1\sin\phi_2 \\
    -z^* \sin\phi_1\sin\phi_2 & 1
  \end{array}
  \right).
\end{eqnarray}
Substituting these matrices into equation (\ref{eqn:eqns-of-motion}),
we find that the equations of motion for $c_1$ and $c_2$ are
\begin{eqnarray}
  \label{eqn:eqns-of-motion-two-level}
  \left(
  \begin{array}{c}
    \dot{c}_1 \\
    \dot{c}_2
  \end{array}
  \right) =
  (1 - |z|^2 \sin^2\phi_1\sin^2\phi_2)^{-1}
  \left(
  \begin{array}{cc}
    |z|^2 \sin\phi_1\cos\phi_1\sin^2\phi_2 & -z \sin\phi_1\cos\phi_2 \\
    -z^* \cos\phi_1\sin\phi_2 & |z|^2 \sin\phi_2\cos\phi_2\sin^2\phi_1
  \end{array}
  \right)
  \left(
  \begin{array}{c}
    \dot{\phi}_1\,c_1 \\
    \dot{\phi}_2\,c_2
  \end{array}
  \right),
\end{eqnarray}
and from equation (\ref{eqn:phi-j}) we have that
\begin{eqnarray}
  \cos \phi_k & = &
  (\cos^2\theta + \lambda_k^2\sin^2\theta)^{-1/2}\,\cos\theta \\
  \sin \phi_k & = &
  (\cos^2\theta + \lambda_k^2\sin^2\theta)^{-1/2}\,\lambda_k \sin\theta \\
  \dot{\phi}_k & = &
  \lambda_k\,(\cos^2\theta + \lambda_k^2\sin^2\theta)^{-1}.
\end{eqnarray}
Given $\lambda_1$, $\lambda_2$, and $z$, these equations of motion can
be integrated to obtain the unitary transformation
$U:{\cal H}_a^\lambda \rightarrow {\cal H}_b^\lambda$.
For certain special cases, we can perform the integration analytically
and write down the explicit form of $U$.

First, suppose $|\psi_{b1}\rangle$ and $|\psi_{b2}\rangle$
are orthogonal.
Then $z = 0$ and the equations of motion reduce to
$\dot{c}_1 = \dot{c}_2 = 0$, so the unitary transformation
$U:{\cal H}_a^\lambda \rightarrow
{\cal H}_b^\lambda$ is given by
\begin{eqnarray}
  U|\psi_{a1}\rangle & = & |\psi_{b2}\rangle \\
  U|\psi_{a2}\rangle & = & |\psi_{b2}\rangle.
\end{eqnarray}

Next, suppose $\lambda_1 = \lambda_2$, so there is a single angle
$\phi = \phi_1 = \phi_2$ that characterizes both lambda dark states.
Also, for simplicity, assume that $z$ is real.
Then we can express the equations of motion as
\begin{eqnarray}
  \label{eqn:eqns-of-motion-simplified}
  \left(
  \begin{array}{c}
    c_1' \\
    c_2'
  \end{array}
  \right) =
  z\,\sin\phi\cos\phi\,(1 - z^2 \sin^4\phi)^{-1}
  \left(
  \begin{array}{cc}
    z \sin^2\phi & -1 \\
    -1 & z \sin^2\phi
  \end{array}
  \right)
  \left(
  \begin{array}{c}
    c_1 \\
    c_2
  \end{array}
  \right),
\end{eqnarray}
where primes denote derivatives with respect to $\phi$.
We can decouple these equations by defining new variables
$\eta_\pm = c_1 \pm c_2$:
\begin{eqnarray}
  \eta_\pm' =
  \mp z \sin\phi \cos\phi\,(1 \pm z\,\sin^2\phi)^{-1}\,\eta_{\pm}.
\end{eqnarray}
These equations can be integrated to give
\begin{eqnarray}
  \label{eqn:eta-integrated}
  \eta_\pm(\phi) =
  (1 \pm z\,\sin^2\phi)^{-1/2}\,\eta_{\pm}(0).
\end{eqnarray}
Thus,
\begin{eqnarray}
  \label{eqn:phi-integrated}
  c_1(\phi) & = &
  \frac{1}{2}[(1 + z\,\sin^2\phi)^{-1/2}\,(c_1(0) + c_2(0)) +
    (1 - z\,\sin^2\phi)^{-1/2}\,(c_1(0) - c_2(0))] \\
  c_2(\phi) & = &
  \frac{1}{2}[(1 + z\,\sin^2\phi)^{-1/2}\,(c_1(0) + c_2(0)) -
    (1 - z\,\sin^2\phi)^{-1/2}\,(c_1(0) - c_2(0))].
\end{eqnarray}
For adiabatic passage from $a$ to $b$ we sweep $\theta$ from $0$ to
$\pi/2$, and from equation (\ref{eqn:phi-j}) we see that $\phi$
also sweeps from $0$ to $\pi/2$.
Thus, after the adiabatic passage has been completed $\phi = \pi/2$,
and the unitary transformation
$U:{\cal H}_a^\lambda \rightarrow {\cal H}_b^\lambda$ is given by
\begin{eqnarray}
  U|\psi_{a1}\rangle & = &
  \alpha\,|\psi_{b1}\rangle + \beta\,|\psi_{b2}\rangle \\
  U|\psi_{a2}\rangle & = &
  \beta\,|\psi_{b1}\rangle + \alpha\,|\psi_{b2}\rangle,
\end{eqnarray}
where
\begin{eqnarray}
  \label{eqn:alpha}
  \alpha & = & \frac{1}{2}((1 + z)^{-1/2} + (1 - z)^{-1/2}) \\
  \label{eqn:beta}
  \beta & = & \frac{1}{2}((1 + z)^{-1/2} - (1 - z)^{-1/2}).
\end{eqnarray}

\section{Example system}
\label{sec:example-system}

\begin{figure}
  \centering
  \includegraphics{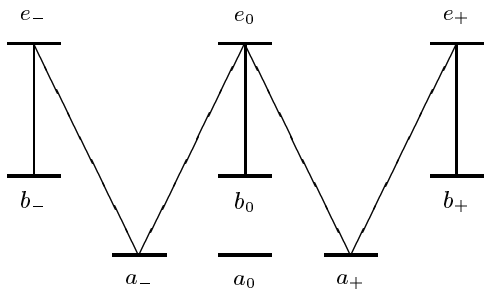}
  \caption{
    \label{fig:example}
    Level diagram for the example system.
    Horizontal lines indicate states in ground state manifolds $a$,
    $b$ and excited state manifold $e$;
    diagonal lines indicate transitions coupled by $A_a$ and
    $A_a^\dagger$; vertical lines indicate transitions coupled by
    $A_b$ and $A_b^\dagger$.
  }
\end{figure}

We will now illustrate the formalism developed in sections
\ref{sec:decompositions} and
\ref{sec:unitary-transformation} with a simple example.
Consider an atom that has the following internal states:
\begin{eqnarray}
  {\cal H}_a & = &
  {\mathrm{span}} \{ |a_+\rangle,\, |a_0\rangle,\, |a_-\rangle \} \\
  {\cal H}_b & = &
  {\mathrm{span}} \{ |b_+\rangle,\, |b_0\rangle,\, |b_-\rangle \} \\
  {\cal H}_e & = &
  {\mathrm{span}} \{ |e_+\rangle,\, |e_0\rangle,\, |e_-\rangle \}.
\end{eqnarray}
Define atomic lowering operators $A_a$ and $A_b$ by
\begin{eqnarray}
  A_b & = &
  |b_+\rangle \langle e_+| +
  |b_0\rangle \langle e_0| +
  |b_-\rangle \langle e_-| \\
  A_a & = &
  |a_+\rangle (\langle e_+| + \langle e_0|) +
  |a_-\rangle (\langle e_-| + \langle e_0|).
\end{eqnarray}
The transitions coupled by these operators are shown in Figure
\ref{fig:example}.

We will first apply the results of section \ref{sec:decompositions} to
find the decompositions of Hilbert spaces ${\cal H}_a$ and
${\cal H}_b$.
There is a single dark state $|a_0\rangle$ for the $a \rightarrow e$
transition, and there are no dark states for the $b \rightarrow e$ and
$e \rightarrow b$ transitions, so
\begin{eqnarray}
  {\cal H}_a^d & = & {\mathrm{span}} \{ |a_0\rangle \} \\
  {\cal H}_a^\perp & = & {\mathrm{span}} \{ |a_\pm \rangle \} \\
  {\cal H}_b^d & = & \{ \} \\
  {\cal H}_b^\perp & = & {\cal H}_b \\
  {\cal H}_e^d & = & \{ \} \\
  {\cal H}_e^\perp & = & {\cal H}_e.
\end{eqnarray}
Since $A_a^\dagger|\psi_a\rangle \in {\cal H}_e^\perp$ for every state
$|\psi_a\rangle \in {\cal H}_a^\perp$, we find that
\begin{eqnarray}
  {\cal H}_a^\lambda & = & {\cal H}_a^\perp =
  {\mathrm{span}} \{ |a_\pm \rangle \} \\
  {\cal H}_a' & = & \{ \}.
\end{eqnarray}
The mapping $M:{\cal H}_a^\lambda \rightarrow {\cal H}_b^\perp$ is
given by
\begin{eqnarray}
  M =
  (|b_+\rangle + |b_0\rangle)\langle a_+| +
  (|b_-\rangle + |b_0\rangle)\langle a_-|.
\end{eqnarray}
We can use this mapping to define states $|B_\pm\rangle$ that form
lambda pairs with states $|a_\pm\rangle$:
\begin{eqnarray}
  |B_\pm\rangle =
  \frac{1}{\lambda_\pm}M|a_\pm\rangle =
  (1/\sqrt{2})(|b_0\rangle + |b_\pm\rangle),
\end{eqnarray}
where
\begin{eqnarray}
  \lambda_\pm =
  \langle a_\pm |M^\dagger M | a_\pm \rangle^{1/2} =
  \sqrt{2}.
\end{eqnarray}
Note that $|B_+\rangle$ and $|B_-\rangle$ are not orthogonal:
\begin{eqnarray}
  z = \langle B_+ | B_- \rangle = 1/2.
\end{eqnarray}
From the results of section \ref{sec:decompositions}, it follows that
\begin{eqnarray}
  |\Lambda_\pm\rangle =
  \cos\phi(\theta)\,|a_\pm\rangle + \sin\phi(\theta)\,|B_\pm\rangle
\end{eqnarray}
are lambda dark states of $H$, where
\begin{eqnarray}
  \label{eqn:phi-example}
  \phi(\theta) = \tan^{-1}(\sqrt{2} \tan \theta).
\end{eqnarray}
Define a state $|s\rangle$ in ${\cal H}_b$ that is orthogonal
to both $|B_+\rangle$ and $|B_-\rangle$:
\begin{eqnarray}
  |s\rangle =
  (1/\sqrt{3})(|b_-\rangle - |b_0\rangle + |b_+\rangle).
\end{eqnarray}
Since the image of $M$ is ${\mathrm{span}} \{ |B_\pm \rangle \}$,
we find that
\begin{eqnarray}
  {\cal H}_b^\lambda & = & {\mathrm{span}} \{ |B_\pm \rangle \} \\
  {\cal H}_b' & = & {\mathrm{span}} \{ |s\rangle \}.
\end{eqnarray}

Now that we have decomposed the Hilbert spaces
${\cal H}_a$ and ${\cal H}_b$, let us apply the results of section
\ref{sec:unitary-transformation} to calculate the unitary
transformation $U:{\cal H}_a^\lambda \rightarrow {\cal H}_b^\lambda$.
We first note that a general lambda dark state of $H$ is a
superposition of $|\Lambda_+\rangle$ and $|\Lambda_-\rangle$:
\begin{eqnarray}
  |\Lambda\rangle = c_1 |\Lambda_+\rangle + c_2 |\Lambda_-\rangle.
\end{eqnarray}
Since $\lambda_+ = \lambda_-$, the equations of motion for $c_1$ and
$c_2$ are given by equation (\ref{eqn:eqns-of-motion-simplified}).
As was shown in section \ref{sec:unitary-transformation-two-level},
these equations of motion can be integrated to give
$U:{\cal H}_a^\lambda \rightarrow {\cal H}_b^\lambda$:
\begin{eqnarray}
  U|a_+\rangle & = &
  \alpha\,|B_+\rangle + \beta\,|B_-\rangle \\
  U|a_-\rangle & = &
  \beta\,|B_+\rangle + \alpha\,|B_-\rangle,
\end{eqnarray}
where $\alpha$ and $\beta$ are given by equations (\ref{eqn:alpha})
and (\ref{eqn:beta}) with $z=1/2$.

The Hilbert space decompositions and unitary transformation $U$ give a
complete description of the behavior of the example system under
adiabatic passage:
for adiabatic passage from $a$ to $b$, states
$|a_\pm\rangle$ coherently evolve into states $U|a_\pm\rangle$ and
state $|a_0\rangle$ remains unchanged;
for adiabatic passage from $b$ to $a$, states
$U|a_\pm\rangle$ coherently evolve into states $|a_\pm\rangle$, and
state $|s\rangle$ is driven to the excited state manifold and scatters
incoherently.

It is interesting to note that if we start in state $|a_+\rangle$ and
perform adiabatic passage from $a$ to $b$, population is transferred
to $|b_-\rangle$ despite the fact that
$\langle b_- |A_b\,A_a^\dagger | a_+ \rangle = 0$.
The population reaches state $|b_-\rangle$ by passing through
state $|a_-\rangle$; at time $t$ the population in $|a_-\rangle$
is given by
\begin{eqnarray}
  P(|a_-\rangle) = |c_2(\phi)|^2\,\cos^2\phi,
\end{eqnarray}
where $c_2(\phi)$ is given by equation (\ref{eqn:phi-integrated}) with 
$c_1(0) = 1$, $c_2(0) = 0$, and where
$\phi = \tan^{-1}(\sqrt{2} \tan \theta(t))$.

\section{Adiabatic passage in Cesium}
\label{sec:cesium}

\begin{figure}
  \centering
  \includegraphics{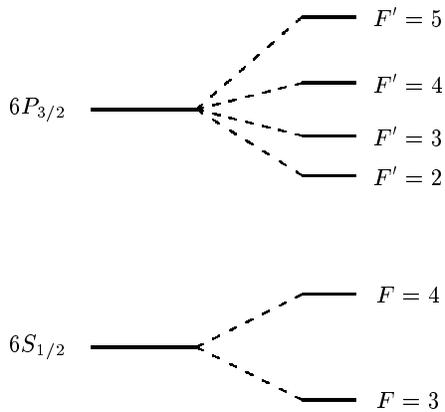}
  \caption{
    \label{fig:cesium-spectrum}
    Level diagram for Cesium.
    The two hyperfine ground state manifolds
    $6S_{1/2}, F=3$ and $6S_{1/2}, F=4$ correspond to manifolds $a$
    and $b$ in our theory, and one of the excited state hyperfine
    manifolds, either $6P_{3/2}, F'=3$ or $6P_{3/2}, F'=4$,
    corresponds to manifold $e$.
    }
\end{figure}

In order to show how the formalism developed in sections
\ref{sec:decompositions} and
\ref{sec:unitary-transformation} applies in a more
physical context, we will use it to analyze adiabatic passage
in the Cesium atom.
A level diagram for Cesium is shown in Figure
\ref{fig:cesium-spectrum};
the two hyperfine ground state manifolds
$6S_{1/2}, F=3$ and $6S_{1/2}, F=4$ correspond to manifolds $a$ and
$b$, and one of the excited state hyperfine
manifolds, either $6P_{3/2}, F'=3$ or $6P_{3/2}, F'=4$, corresponds
to manifold $e$.
For simplicity, we will denote the
$6S_{1/2}, F=3$ and $6S_{1/2}, F=4$
ground state manifolds by $g_3$ and $g_4$, and the
$6P_{3/2}, F'=3$ and $6P_{3/2}, F'=4$ excited state manifolds by $e_3$
and $e_4$.

Adiabatic passage between manifolds $g_3$ and $g_4$ is performed by
driving the atom with a pair of classical fields that connect these
manifolds to the excited state manifold $e_{F'}$.
The field driving the $g_F \leftrightarrow e_{F'}$ transition
corresponds to an atomic lowering operator that is given by
\begin{eqnarray}
  A_F = \vec{A}_F\cdot\hat{\epsilon}_F,
\end{eqnarray}
where $\hat{\epsilon}_F$ is the polarization of the field.
The operator $\vec{A}_F$ is defined by
\begin{eqnarray}
  \label{eqn:vector-rasing-operator}
  \vec{A}_F^\dagger \equiv
  \sum_{m'}\sum_m \sum_{q=-1}^1 \beta(F',F)\,
  \langle F',m'|1,q;F,m\rangle\,|F',m'\rangle \langle F,m|\,
  \hat{e}_q^*,
\end{eqnarray}
where $\langle F',m'|1,q;F,m\rangle$ is the Clebsch-Gordan coefficient
that connects ground state $|F,m\rangle$ to excited state
$|F',m'\rangle$ via polarization $\hat{e}_q^*$,
\begin{eqnarray}
  \label{eqn:circular-polarization-vectors}
  \hat{e}_{\pm 1} & = &
  \mp \frac{1}{\sqrt{2}}(\hat{x} \pm i \hat{y}) \\
  \hat{e}_0 & = & \hat{z},
\end{eqnarray}
is a orthonormal basis of polarization vectors,
and $\beta(F',F)$ is given by the following table:
\begin{eqnarray}
  \begin{array}{|c|cccc|}
    \hline
    F' & 3 & 3 & 4 & 4 \\
    F & 3 & 4 & 3 & 4 \\
    \beta(F',F) & \sqrt{3/4} & \sqrt{1/4} & \sqrt{5/12} & \sqrt{7/12} \\
    \hline
  \end{array}
\end{eqnarray}
In the following sections we apply the results of section
\ref{sec:decompositions} to perform the Hilbert space decompositions
given in equations
(\ref{eqn:decomposition-a}) and
(\ref{eqn:decomposition-b})
for adiabatic passage via both the $F'=3$ and $F'=4$ excited state
manifolds.

\subsection{Adiabatic passage via the $F'=3$ manifold}
\label{ssec:passage-via-e3}

For the $g_4 \leftrightarrow e_3$ transition there are two dark states
$|d_{4e}\rangle$, $|d_{4o}\rangle$ in $g_4$ (these
dark states are calculated in Appendix
\ref{dark_state_f_to_f_minus_1}),
and there are no dark states in $e_3$, so
\begin{eqnarray}
  {\cal H}_4^d & = & {\mathrm{span}}
  \{ |d_{4e}\rangle,\, |d_{4o}\rangle \} \\
  {\cal H}_e^d & = & \{ \}.
\end{eqnarray}
For the $g_3 \leftrightarrow e_3$ transition there is a single dark
state $|d_3\rangle$ in $g_3$ (this dark state is calculated in Appendix
\ref{dark_state_f_to_f}), so
\begin{eqnarray}
  {\cal H}_3^d & = & {\mathrm{span}} \{ |d_3\rangle \}.
\end{eqnarray}
From the dimensions of ${\cal H}_3^d$, ${\cal H}_4^d$, and
${\cal H}_e^d$, we can work out the dimensions of all the Hilbert
spaces in the decompositions of ${\cal H}_3$ and
${\cal H}_4$.
First, note that because $\dim {\cal H}_e^d = 0$ and
$\dim {\cal H}_3' \leq \dim {\cal H}_e^d$, we must have that
$\dim {\cal H}_3' = 0$.
Next, since the dimensions of
${\cal H}_3^\lambda$, ${\cal H}_3^d$, and ${\cal H}_3'$ must sum to
$7$, we find that $\dim {\cal H}_3^\lambda = 6$.
Finally, since $\dim {\cal H}_4^\lambda = \dim {\cal H}_3^\lambda = 6$
and the dimensions of
${\cal H}_4^\lambda$, ${\cal H}_4^d$, and ${\cal H}_4'$ must sum
to $9$, we find that $\dim {\cal H}_4' = 1$.
Thus, the dimensions of the Hilbert spaces are
\begin{eqnarray*}
  \begin{tabular}{|c|ccc|}
    \hline
    F & $\dim {\cal H}_F^\lambda $ &  $\dim {\cal H}_F^d$ &
    $\dim {\cal H}_F'$ \\
    \hline
    3 &  6 & 1 & 0  \\
    4 &  6 & 2 & 1 \\
    \hline
  \end{tabular}
\end{eqnarray*}
Note that these dimensions are independent of the polarizations of the
classical fields.

\subsubsection{Example:
$\hat{\epsilon}_3 = \hat{\epsilon}_4 = \hat{z}$}

The Hilbert space decompositions are
\begin{eqnarray}
  {\cal H}_3^d & = &
  {\mathrm{span}} \{ |3,0\rangle \} \\
  {\cal H}_3' & = &
  \{  \} \\
  {\cal H}_3^\lambda & = &
  {\mathrm{span}} \{ |3,\pm 1\rangle,\, |3, \pm 2\rangle,\,
  |3, \pm 3\rangle \} \\
  {\cal H}_4^d & = &
  {\mathrm{span}} \{ |4, \pm 4\rangle \} \\
  {\cal H}_4' & = &
  {\mathrm{span}} \{ |4, 0\rangle \} \\
  {\cal H}_4^\lambda & = &
  {\mathrm{span}} \{ |4,\pm 1\rangle,\, |4, \pm 2\rangle,\,
  |4, \pm 3\rangle \}.
\end{eqnarray}
States $|3,m\rangle$ and $|4,m\rangle$ are lambda pairs for
$m=\pm 1,\pm 2, \pm 3$.
Since the lambda dark states corresponding to these pairs are mutually
orthogonal, equation (\ref{eqn:eqns-of-motion}) decouples and the
unitary transformation
$U:{\cal H}_3^\lambda \rightarrow {\cal H}_4^\lambda$ can be written
down explicitly:
\begin{eqnarray}
  U = \sum_m |4,m\rangle \langle 3,m|,
\end{eqnarray}
where the sum is taken over $m=\pm 1,\pm 2, \pm 3$.

\subsubsection{Example:
$\hat{\epsilon}_3 = \hat{z}$, $\hat{\epsilon}_4 = \hat{x}$}
\label{ssec:example-3-2}

The Hilbert space decompositions are
\begin{eqnarray}
  {\cal H}_3^d & = &
  {\mathrm{span}} \{ |3,0\rangle \} \\
  {\cal H}_3' & = &
  \{  \} \\
  {\cal H}_3^\lambda & = &
  {\mathrm{span}} \{ |3,\pm 1\rangle,\, |3, \pm 2\rangle,\,
  |3, \pm 3\rangle \} \\
  {\cal H}_4^d & = &
  {\mathrm{span}} \{ |d_{4e}\rangle,\, |d_{4o}\rangle \} \\
  {\cal H}_4' & = &
  {\mathrm{span}} \{ |s\rangle \} \\
  {\cal H}_4^\lambda & = &
  {\mathrm{span}} \{ |B_{\pm 1}\rangle,\,
  |B_{\pm 2}\rangle,\,
  |B_{\pm 3}\rangle \},
\end{eqnarray}
where
\begin{eqnarray}
  |s\rangle \equiv (1/\sqrt{2})\,(|4,1\rangle - |4,-1\rangle)
\end{eqnarray}
and the dark states are given by
\begin{eqnarray}
  |d_{4e}\rangle & = &
  (1/8\sqrt{2})\,(
  |4,4\rangle +
  \sqrt{28}\, |4,2\rangle +
  \sqrt{70}\, |4,0\rangle +
  \sqrt{28}\, |4,-2\rangle +
  |4,-4\rangle) \\
  |d_{4o}\rangle & = &
  (1/4)\,(
  |4,3\rangle +
  \sqrt{7}\, |4,1\rangle +
  \sqrt{7}\, |4,-1\rangle +
  |4,-3\rangle).
\end{eqnarray}
We have defined states $|B_{\pm 1}\rangle$, $|B_{\pm 2}\rangle$, and
$|B_{\pm 3}\rangle$ that form lambda pairs with states
$|3,\pm 1\rangle$, $|3,\pm 2\rangle$, and $|3,\pm 3\rangle$; they are
given by
\begin{eqnarray}
  |B_{\pm 3}\rangle & = &
  (1/8\sqrt{254})\,(
  127 |4,\pm 4\rangle -
  2\sqrt{7} |4, \pm2\rangle -
  \sqrt{70} |4, 0\rangle -
  2\sqrt{7} |4, \mp 2\rangle -
  |4,\mp 4\rangle ) \\
  |B_{\pm 2}\rangle & = &
  (1/4\sqrt{15})\,(
  15 |4,\pm 3\rangle -
  \sqrt{7} |4, \pm1\rangle -
  \sqrt{7} |4, \mp 1\rangle -
  |4,\mp 3\rangle ) \\
  |B_{\pm 1}\rangle & = &
  (1/24\sqrt{638})\,(
  99\,|4,\pm 4\rangle +
  198 \sqrt{7} |4, \pm2\rangle -
  29 \sqrt{70} |4, 0\rangle -
  58 \sqrt{7} |4, \mp 2\rangle -
  29 |4,\mp 4\rangle ),
\end{eqnarray}
and the corresponding $\lambda$ values are
\begin{eqnarray}
  \lambda_{\pm 3} & = & \sqrt{10287/1792} \\
  \lambda_{\pm 2} & = & \sqrt{45/14} \\
  \lambda_{\pm 1} & = & \sqrt{8613/8960}.
\end{eqnarray}
The unitary transformation
$U:{\cal H}_3^\lambda \rightarrow {\cal H}_4^\lambda$ can be obtained
by numerically integrating equation (\ref{eqn:eqns-of-motion}) using
the above expressions for the states and $\lambda$ values.
We can say something about the structure of this unitary
transformation by noting that the polarizations of the classical
fields impose the selection rule $\Delta m = \pm 1$.
Thus, it is useful to decompose
${\cal H}_3^\lambda$ and ${\cal H}_4^\lambda$ into subspaces of even
and odd Zeeman states:
\begin{eqnarray}
  {\cal H}_3^\lambda & = &
  {\cal H}_3^{\lambda e} \oplus {\cal H}_3^{\lambda o} \\
  {\cal H}_4^\lambda & = &
  {\cal H}_4^{\lambda e} \oplus {\cal H}_4^{\lambda o},
\end{eqnarray}
where
\begin{eqnarray}
  {\cal H}_3^{\lambda e} & = &
  {\mathrm{span}} \{ |3, \pm 2\rangle \} \\
  {\cal H}_3^{\lambda o} & = &
  {\mathrm{span}} \{ |3,\pm 1\rangle,\, |3, \pm 3\rangle \} \\
  {\cal H}_4^{\lambda o} & = &
  {\mathrm{span}} \{ |B_{\pm 2} \rangle \} \\
  {\cal H}_4^{\lambda e} & = &
  {\mathrm{span}} \{ |B_{\pm 1}\rangle,\, |B_{\pm 3}\rangle \}.
\end{eqnarray}
The selection rule implies that U maps
${\cal H}_3^{\lambda e}$ to ${\cal H}_4^{\lambda o}$ and
${\cal H}_3^{\lambda o}$ to ${\cal H}_4^{\lambda e}$.
Since
$\dim {\cal H}_3^{\lambda e} = \dim {\cal H}_4^{\lambda o} = 2$ and
$\lambda_{+2} = \lambda_{-2}$, we can use the results of section
\ref{sec:unitary-transformation-two-level} to
write down the unitary transformation
$U:{\cal H}_3^{\lambda e} \rightarrow {\cal H}_4^{\lambda o}$
explicitly:
\begin{eqnarray}
  U|3,+2\rangle & = &
  \alpha\,|B_{+2}\rangle + \beta\,|B_{-2}\rangle \\
  U|3,-2\rangle & = &
  \beta\,|B_{+2}\rangle + \alpha\,|B_{-2}\rangle,
\end{eqnarray}
where $\alpha$ and $\beta$ are given by equations
(\ref{eqn:alpha}) and (\ref{eqn:beta}) with
$z = \langle B_{+2} | B_{-2} \rangle = -1/15$.

\subsection{Adiabatic passage via the $F'=4$ manifold}
\label{ssec:passage-via-e4}

For the $g_4 \leftrightarrow e_4$ transition there is one dark state
$|d_4\rangle$ in $g_4$ and one dark state $|d_e\rangle$ in $e_4$
(these dark states are calculated in Appendix \ref{dark_state_f_to_f}),
so
\begin{eqnarray}
  {\cal H}_4^d & = & {\mathrm{span}} \{ |d_4\rangle \} \\
  {\cal H}_e^d & = & {\mathrm{span}} \{ |d_e\rangle \}.
\end{eqnarray}
For the $g_3 \leftrightarrow e_4$ transition there are no dark states
in $g_3$, so
\begin{eqnarray}
  {\cal H}_3^d & = & \{ \}.
\end{eqnarray}
As with adiabatic passage via $F'=3$, we can use the dimensions of
${\cal H}_3^d$, ${\cal H}_4^d$, and
${\cal H}_e^d$ to say something about the dimensions of the other
Hilbert spaces in the decomposition of ${\cal H}_3$ and
${\cal H}_4$.
For adiabatic passage via $F'=4$, however, there are two separate
cases to consider: since $\dim {\cal H}_3' \leq \dim{\cal H}_e^d$ and
$\dim{\cal H}_e^d = 1$, we find that $\dim {\cal H}_3'$ can be either
$0$ or $1$.
For $\dim {\cal H}_3' = 0$ the dimensions of the Hilbert spaces are
\begin{eqnarray*}
\begin{tabular}{|c|ccc|}
  \hline
  F & $\dim {\cal H}_F^\lambda $ &  $\dim {\cal H}_F^d$ &
  $\dim {\cal H}_F'$ \\
  \hline
  3 &  7 & 0 & 0  \\
  4 &  7 & 1 & 1 \\
  \hline
\end{tabular}
\end{eqnarray*}
and for $\dim {\cal H}_3' = 1$ the dimensions of the Hilbert spaces
are
\begin{eqnarray*}
\begin{tabular}{|c|ccc|}
  \hline
  F & $\dim {\cal H}_F^\lambda $ &  $\dim {\cal H}_F^d$ &
  $\dim {\cal H}_F'$ \\
  \hline
  3 &  6 & 0 & 1  \\
  4 &  6 & 1 & 2 \\
  \hline
\end{tabular}
\end{eqnarray*}
Thus, for adiabatic passage via $F'=4$ the dimensions of the Hilbert
spaces depend on the polarizations of the classical fields.

\subsubsection{Example:
$\hat{\epsilon}_3 = \hat{\epsilon}_4 = \hat{z}$}

For this example $\dim {\cal H}_3' = 1$, and the Hilbert space
decompositions are
\begin{eqnarray}
  {\cal H}_3^d & = &
  \{ \} \\
  {\cal H}_3' & = &
  {\mathrm{span}} \{ |3, 0\rangle \} \\
  {\cal H}_3^\lambda & = &
  {\mathrm{span}} \{ |3,\pm 1\rangle,\, |3, \pm 2\rangle,\,
  |3, \pm 3\rangle \} \\
  {\cal H}_4^d & = &
  {\mathrm{span}} \{ |4, 0\rangle \} \\
  {\cal H}_4' & = &
  {\mathrm{span}} \{ |4, \pm 4\rangle \} \\
  {\cal H}_4^\lambda & = &
  {\mathrm{span}} \{ |4,\pm 1\rangle,\, |4, \pm 2\rangle,\,
  |4, \pm 3\rangle \}.
\end{eqnarray}
States $|3,m\rangle$ and $|4,m\rangle$ are lambda pairs for
$m=\pm 1,\pm 2, \pm 3$.
Since the lambda dark states corresponding to these pairs are mutually
orthogonal, equation (\ref{eqn:eqns-of-motion}) decouples, and the
unitary transformation
$U:{\cal H}_3^\lambda \rightarrow {\cal H}_4^\lambda$ can be written
down explicitly:
\begin{eqnarray}
  U = \sum_m |4,m\rangle \langle 3,m|,
\end{eqnarray}
where the sum is taken over $m=\pm 1,\pm 2, \pm 3$.

\subsubsection{Example:
$\hat{\epsilon}_3 = \hat{x}$, $\hat{\epsilon}_4 = \hat{z}$}

For this example $\dim {\cal H}_3' = 1$, and the Hilbert space
decompositions are
\begin{eqnarray}
  {\cal H}_3^d & = &
  \{ \} \\
  {\cal H}_3' & = &
  {\mathrm{span}} \{  |-\rangle \} \\
  {\cal H}_3^\lambda & = &
  {\mathrm{span}} \{ |3,\pm 2\rangle,\, |3, \pm 3\rangle,\,
  |3, 0\rangle,\, |+\rangle \} \\
  {\cal H}_4^d & = &
  {\mathrm{span}} \{ |4,0\rangle \} \\
  {\cal H}_4' & = &
  {\mathrm{span}} \{ |s_e\rangle,\, |s_o\rangle \} \\
  {\cal H}_4^\lambda & = &
  {\mathrm{span}} \{ |B_{\pm 2} \rangle,\,
  |B_{\pm 3} \rangle,\, |B_0\rangle,\, |B_+\rangle \},
\end{eqnarray}
where
\begin{eqnarray}
  |\pm\rangle \equiv (1/\sqrt{2})(|3,1\rangle \pm |3,-1\rangle)
\end{eqnarray}
and
\begin{eqnarray}
  |s_e \rangle & \equiv &
  (1/4)\,(|4,4\rangle + \sqrt{7}|4,2\rangle -
  \sqrt{7}|4,-2\rangle - |4,-4\rangle) \\
  |s_o \rangle & \equiv &
  (1/\sqrt{32})\,(
  3|4,3\rangle + \sqrt{7}|4,1\rangle -
  \sqrt{7}|4,-1\rangle - 3|4,-3\rangle).
\end{eqnarray}
We have defined states $|B_+\rangle$, $|B_0\rangle$,
$|B_{\pm 2}\rangle$, and $|B_{\pm 3}\rangle$ that form lambda pairs
with states
$|+\rangle$, $|3,0\rangle$, $|3,\pm 2\rangle$, and $|3,\pm 3\rangle$;
they are given by
\begin{eqnarray}
  |B_{\pm 3} \rangle & = &
  (1/\sqrt{8})\,(\sqrt{7}|4, \pm4\rangle - |4, \pm2\rangle) \\
  |B_{\pm 2} \rangle & = &
  (1/4)(\sqrt{7}|4, \pm3\rangle - 3|4, \pm1\rangle) \\
  |B_0\rangle & = &
  (1/\sqrt{2})\,(|4,1\rangle + |4,-1\rangle) \\
  |B_+ \rangle & = &
  (1/\sqrt{2})\,(|4,2\rangle + |4,-2\rangle),
\end{eqnarray}
and the corresponding $\lambda$ values are
\begin{eqnarray}
  \lambda_{\pm 3} & = & \sqrt{25/49} \\
  \lambda_{\pm 2} & = & \sqrt{200/147} \\
  \lambda_0 & = & \sqrt{250/49} \\
  \lambda_+ & = & \sqrt{375/392}.
\end{eqnarray}
The unitary transformation
$U:{\cal H}_3^\lambda \rightarrow {\cal H}_4^\lambda$ can be obtained
by numerically integrating equation (\ref{eqn:eqns-of-motion}) using
the above expressions for the states and $\lambda$ values.
As with the example given in section \ref{ssec:example-3-2}, the
polarizations of the classical fields impose the selection rule
$\Delta m = \pm 1$, so it is useful to decompose ${\cal H}_3^\lambda$
and ${\cal H}_4^\lambda$ into subspaces of even and odd Zeeman states:
\begin{eqnarray}
  {\cal H}_3^\lambda & = &
  {\cal H}_3^{\lambda e} \oplus {\cal H}_3^{\lambda o} \\
  {\cal H}_4^\lambda & = &
  {\cal H}_4^{\lambda e} \oplus {\cal H}_4^{\lambda o},
\end{eqnarray}
where
\begin{eqnarray}
  {\cal H}_3^{\lambda e} & = &
  {\mathrm{span}} \{ |3,0\rangle,\, |3, \pm 2\rangle \} \\
  {\cal H}_3^{\lambda o} & = &
  {\mathrm{span}} \{ |+\rangle,\, |3, \pm 3\rangle \} \\
  {\cal H}_4^{\lambda o} & = &
  {\mathrm{span}} \{ |B_0\rangle,\, |B_{\pm 2} \rangle \} \\
  {\cal H}_4^{\lambda e} & = &
  {\mathrm{span}} \{ |B_+\rangle,\, |B_{\pm 3}\rangle \}.
\end{eqnarray}
The selection rule implies that $U$ maps
${\cal H}_3^{\lambda e}$ to ${\cal H}_4^{\lambda o}$ and
${\cal H}_3^{\lambda o}$ to ${\cal H}_4^{\lambda e}$.

\subsubsection{Example:
$\hat{\epsilon}_3 = \hat{\epsilon}_4 = \hat{e}_{+1}$}

For this example $\dim {\cal H}_3' = 0$, and the Hilbert space
decompositions are
\begin{eqnarray}
  {\cal H}_3^d & = &
  \{ \} \\
  {\cal H}_3' & = &
  {\mathrm{span}} \{  \} \\
  {\cal H}_3^\lambda & = &
  {\mathrm{span}} \{ |3,0\rangle,\, |3,\pm 1\rangle,\,
  |3, \pm 2\rangle,\, |3, \pm 3\rangle \} \\
  {\cal H}_4^d & = &
  {\mathrm{span}} \{ |4, 4\rangle \} \\
  {\cal H}_4' & = &
  {\mathrm{span}} \{ |4, -4\rangle \} \\
  {\cal H}_4^\lambda & = &
  {\mathrm{span}} \{ |4,0\rangle,\, |4,\pm 1\rangle,\,
  |4, \pm 2\rangle,\, |4, \pm 3\rangle \}.
\end{eqnarray}
States $|3,m\rangle$ and $|4,m\rangle$ are lambda pairs for
$m=0,\pm 1,\pm 2, \pm 3$.
Since the lambda dark states corresponding to these pairs are mutually
orthogonal, equation (\ref{eqn:eqns-of-motion}) decouples, and the
unitary transformation
$U:{\cal H}_3^\lambda \rightarrow {\cal H}_4^\lambda$ can be written
down explicitly:
\begin{eqnarray}
  U = \sum_{m=-3}^3 |4,m\rangle \langle 3,m|.
\end{eqnarray}

\section{Conclusion}

We have presented a general formalism for describing stimulated
Raman adiabatic passage in a multi-level atom with ground state
manifolds $a$ and $b$ and excited state manifold $e$, where each
manifold consists of an arbitrary number of degenerate states.
This formalism describes how the Hilbert spaces for manifolds $a$ and
$b$ decompose into subspaces, each of which evolves in its own
characteristic way under adiabatic passage.
In particular, we identified subspaces ${\cal H}_a^\lambda$ and
${\cal H}_b^\lambda$ that coherently evolve into one another under
adiabatic passage, and calculated the unitary transformation that
describes this evolution.
We have illustrated the formalism with a simple toy model, and used it
to analyze adiabatic passage in the Cesium atom.
The formalism gives a complete description of the adiabatic passage
process, and should be useful for analyzing adiabatic passage in a
wide variety of atomic systems.

\acknowledgements

The author would like to thank A. Boca and T. E. Northup for helpful
suggestions.

\appendix

\section{Dark states}

Here we calculate the dark states for an optical field driving an 
$F \rightarrow F$ or $F \rightarrow F-1$ transition.
We will choose a coordinate system such that the field propagates
along the $\hat{z}$ axis, and define a basis of Zeeman states relative
to this quantization axis.
The polarization $\hat{\epsilon}$ of the field can be expressed as
\begin{eqnarray}
  \hat{\epsilon} = \alpha_+ \hat{e}_{+1} + \alpha_- \hat{e}_{-1},
\end{eqnarray}
where $\hat{e}_{\pm 1}$ are given by equation
(\ref{eqn:circular-polarization-vectors}).
If $|d\rangle$ is a dark state of the field, then
\begin{eqnarray}
  \hat{\epsilon}\cdot \vec{A}^\dagger
  |d\rangle = 0.
\end{eqnarray}
Thus, for every state $|F',m'\rangle$ in the excited state manifold,
we have that
\begin{eqnarray}
  \sum_m\,
  \langle F',m' |\hat{\epsilon}\cdot \vec{A}^\dagger |F,m\rangle\,
  c_m = 0,
\end{eqnarray}
where
\begin{eqnarray}
  c_m \equiv \langle F, m|d\rangle.
\end{eqnarray}
Using equation (\ref{eqn:vector-rasing-operator}) to substitute
for $\vec{A}_F^\dagger$, we obtain the following dark state
conditions:
\begin{eqnarray}
  \langle F',m|1,+1; F,m-1\rangle\,\alpha_+ c_{m-1} +
  \langle F',m|1,-1; F,m+1\rangle\,\alpha_- c_{m+1} = 0
\end{eqnarray}
for $-F' \leq m \leq F'$.
We will now consider $F \rightarrow F$ and $F \rightarrow F-1$
transitions as separate cases.

\subsection{Dark state for an $F \rightarrow F$ transition}
\label{dark_state_f_to_f}

The Clebsch-Gordan coefficients for $-F < m < F$ are given by
\begin{eqnarray}
  \langle F,m|1,+1; F,m-1\rangle & = &
  -\left(\frac{(F+m)(F+1-m)}{2F(F+1)}\right)^{1/2} \\
  \langle F,m|1,-1; F,m+1\rangle & = &
  +\left(\frac{(F-m)(F+1+m)}{2F(F+1)}\right)^{1/2}.
\end{eqnarray}
Thus, the dark state conditions are
\begin{eqnarray}
  \sqrt{(F+m)(F+1-m)}\,\alpha_+ c_{m-1} = 
  \sqrt{(F-m)(F+1+m)}\,\alpha_- c_{m+1}
\end{eqnarray}
for $-F < m < F$, and
\begin{eqnarray}
  \alpha_+ c_{F-1} = \alpha_- c_{-(F-1)} = 0.
\end{eqnarray}
For each value of $F$ there is single dark state that meets these
conditions.
The dark state for a $3 \rightarrow 3$ transition is
\begin{eqnarray}
  |d_3\rangle =
  N(
  \sqrt{5}\, \alpha_+^3 |3,3\rangle +
  \sqrt{3}\, \alpha_+^2 \alpha_- |3,1\rangle +
  \sqrt{3}\, \alpha_+ \alpha_-^2 |3,-1\rangle +
  \sqrt{5}\, \alpha_-^3 |3,-3\rangle),
\end{eqnarray}
and the dark state for a $4\rightarrow 4$ transition is
\begin{eqnarray}
  |d_4\rangle =
  N(
  \sqrt{35}\, \alpha_+^4 |4,4\rangle +
  \sqrt{20}\, \alpha_+^3 \alpha_- |4,2\rangle +
  \sqrt{18}\, \alpha_+^2 \alpha_-^2 |4,0\rangle +
  \sqrt{20}\, \alpha_+ \alpha_-^3 |4,-2\rangle +
  \sqrt{35}\, \alpha_-^4 |4,-4\rangle),
\end{eqnarray}
where $N$ is a normalization constant that depends on $\alpha_\pm$.

\subsection{Dark states for an $F \rightarrow F-1$ transition}
\label{dark_state_f_to_f_minus_1}

The Clebsch-Gordan coefficients for $-F < m < F$ are given by
\begin{eqnarray}
  \langle F-1,m|1,+1; F,m-1\rangle & = &
  \left(\frac{(F-m)(F+1-m)}{2F(2F+1)}\right)^{1/2} \\
  \langle F-1,m|1,-1; F,m+1\rangle & = &
  \left(\frac{(F+m)(F+1+m)}{2F(2F+1)}\right)^{1/2}.
\end{eqnarray}
Thus, the dark state conditions are
\begin{eqnarray}
  \sqrt{(F-m)(F+1-m)}\,\alpha_+ c_{m-1} = 
  -\sqrt{(F+m)(F+1+m)}\,\alpha_- c_{m+1}.
\end{eqnarray}
for $-F < m < F$.
For each value of $F$ there are two dark states that meet these
conditions.
The dark states for a $4\rightarrow 3$ transition are
\begin{eqnarray}
  |d_{4e}\rangle & = &
  N(
  \alpha_+^4 |4,4\rangle -
  \sqrt{28}\, \alpha_+^3 \alpha_- |4,2\rangle +
  \sqrt{70}\, \alpha_+^2 \alpha_-^2 |4,0\rangle -
  \sqrt{28}\, \alpha_+ \alpha_-^3 |4,-2\rangle +
  \alpha_-^4 |4,-4\rangle) \\
  |d_{4o}\rangle & = &
  N(
  \alpha_+^3 |4,3\rangle -
  \sqrt{7}\, \alpha_+^2 \alpha_- |4,1\rangle +
  \sqrt{7}\, \alpha_+ \alpha_-^2 |4,-1\rangle -
  \alpha_-^3 |4,-3\rangle),
\end{eqnarray}
where $N$ is a normalization constant that depends on $\alpha_\pm$.


\begin{thebibliography}{99}


\bibitem{hioe83}
  F. T. Hioe,
  Phys. Lett. A \textbf{99}, 150 (1983).

\bibitem{oreg84}
  J. Oreg, F. T. Hioe, and J. H. Eberly,
  Phys. Rev. A \textbf{29}, 690 (1984).

\bibitem{kuklinski89}%
  J. R. Kuklinski, U. Gaubatz, F. T. Hioe, and K. Bergmann,
  Phys. Rev. A \textbf{40}, 6741 (1989).


\bibitem{gaubatz88}
  U. Gaubatz, P. Rudecki, M. Becker, S. Schiemann, M. K\"{u}lz,
  and K. Bergmann,
  Chem. Phys. Lett. \textbf{149}, 463 (1988).

\bibitem{gaubatz90}
  U. Gaubatz, P. Rudecki, S. Schiemann, and K. Bergmann,
  J. Chem. Phys. \textbf{92}, 5363 (1990).

\bibitem{rubahn91}
  H. G. Rubahn, E. Konz, S. Schiemann, and K. Bergmann,
  Z. Phys. D \textbf{22}, 401 (1991).

\bibitem{broers92}%
  B. Broers, H. B. van Linden van den Heuvell, and L. D. Noordam,
  Phys. Rev. Lett. \textbf{69}, 2062 (1992).

\bibitem{schiemann93}%
  S. Schiemann, A. Kuhn, S. Steuerwald, and K. Bergmann,
  Phys. Rev. Lett. \textbf{71}, 3637 (1993).

\bibitem{weitz94}%
  M. Weitz, B. C. Young, and S. Chu,
  Phys. Rev. Lett. \textbf{73} 2563 (1994).

\bibitem{boozer07}
  A. D. Boozer, A. Boca, R. Miller, T. E. Northup, and H. J. Kimble,
  Phys. Rev. Lett. \textbf{98}, 193601 (2007)


\bibitem{marte91}%
  P. Marte, P. Zoller, and J. L. Hall,
  Phys. Rev. A \textbf{44} R4118 (1991)

\bibitem{unanyan98}%
  R. G. Unanyan, M. Fleischhaur, B. W. Shore, and K. Bergmann,
  Opt. Commun. \textbf{155} 144 (1998).

\bibitem{theuer99}%
  H. Theuer, R. Unanyan, C. Habscheid, K. Klein, and K. Bergmann,
   Opt. Express \textbf{4} 77 (1999).

\bibitem{unanyan99}%
  R. G. Unanyan, B. W. Shore, and K. Bergmann,
  Phys. Rev. A \textbf{59} 2910 (1999).


\bibitem{shore95}%
  B. W. Shore, J. Martin, M. P. Fewell, and K. Bergmann
  Phys. Rev. A \textbf{52}, 566 (1995)

\bibitem{martin95}%
  J. Martin, B. W. Shore, K. Bergmann
  Phys. Rev. A \textbf{52}, 583 (1995)

\bibitem{kis01}%
  Z. Kis and S. Stenholm,
  Phys. Rev. A \textbf{64}, 063406 (2001)

\bibitem{kiz04}%
  Z. Kis, A. Karpati, B. W. Shore, and N. V. Vitanov,
  Phys. Rev. A \textbf{70}, 053405 (2004)


\bibitem{chang01}%
  B. Y. Chang, I. R. Sol\'{a}, V. S. Malinovsky, and J. Santamar,
  Phys. Rev. A \textbf{64} 033420 (2001).

\bibitem{kis02}%
  Z. Kis and S. Stenholm,
  J. Mod. Opt. \textbf{49}, 111, (2002)

\bibitem{karpati03}%
  A. Karpati and Z. Kis,
  J. Phys. B \textbf{36}, 905 (2003).

\bibitem{kis05}%
  Z. Kis, N. V. Vitanov, A. Karpati, C. Barthel, and K. Bergmann,
  Phys. Rev. A \textbf{72}, 033403 (2005)


\bibitem{thanopulos}
  I. Thanopulos, P. Kr\'{a}l, and M. Shapiro.
  Phys. Rev. Lett. 92, 113003 (2004)

\end{thebibliography}
\end{document}